\documentclass[aps,prl,twocolumn,groupedaddress]{revtex4}

\usepackage{amssymb}

\usepackage{graphicx}
\usepackage{graphics}
\begin{document}

\title{Large scale shell model calculations along Z=28 and N=50 closures:
towards the doubly-magic $^{78}$Ni}
\author{K.~Sieja and F.~Nowacki }

\address{Institute Pluridisciplinaire Hubert Curien, 23 rue du Loess, 
Strasbourg, France}
\date{today}

\begin{abstract}
We present the state-of-the art shell model calculations
in a large model space ($pf$ for protons, $fpgd$ for neutrons), which allow to study simultaneously excitations
across the $Z=28$ and $N=50$ shell gaps. We explore the region in the vicinity of
$^{78}$Ni, being a subject of intense experimental investigations.
Our calculations account correctly for the known low lying excited states in this region, 
including those which may correspond to cross-shell excitations.
We observe the minimum of the $N=50$ mass gap at $Z=32$ consistent with experimental data 
and its further increase towards $Z=28$, indicating a robustness of the $N=50$
gap in $^{78}$Ni. The evolution of $N=50$ gap along the nickel chain is shown to bear similarities
with what is know in oxygen and calcium chains, providing a new opportunity
for the studies of 3-body monopole effects in medium mass nuclei.

\end{abstract}

\maketitle

The search for the breaking of shell closures
known at the stability valley when going towards the drip lines
is one of the main interests of contemporary nuclear structure studies.
Last decades provided many examples of unexpected shell erosions, $^{42}$Si
being a famous example \cite{Bastin2007}, 
and of appearances of deformed intruders in supposedly semi-magic nuclei,
known as the phenomena of {\it islands of inversions} \cite{Utsuno-island,Lenzi2010}.
The issue of the occurence of shell quenching in connection with 
astrophysical scenarios has been widely debated, in particular for $N=50$
and $N=82$ around the $r$-process waiting points \cite{Dillmann, Jungclaus, Caceres}.

Experimental progress allows for studying more and more exotic systems
including the nuclear structure towards the still unknown, possibly doubly magic
nucleus: $^{78}$Ni. The region around $^{78}$Ni is particulary interesting for several reasons.
As this nucleus (in some scenarios) is one of the waiting-points in the $r$-process,
testing the rigidity of its gaps remains of a special interest. 
Further, the evolution of the $N=50$ gap between $^{68}$Ni and $^{78}$Ni may be due to the repulsive character 
of the effective 3-body force, in analogy to what has been found in
oxygen and calcium chains \cite{Holt1, Holt2}. Thus constraining
its size is of a paramount importance for future developments and tests of      
state-of-the art effective interactions with the inclusion of many-body forces.
Finally, the knowledge of single particle energies of $^{78}$Ni is
crucial for shell model studies which utilise this nucleus as a core \cite{sieja-zr}
and to validate the empirical {\it universal monopole} interactions, like those proposed 
in Refs. \cite{GEMO, Otsuka2006, Otsuka2010}. 

Though a lot of experimental evidence in this region has been accumulated,
the data, or to be precise the conclusions drawn by different authors, 
seem contradictory. The possibility of the weakening of the $N=50$ closure has
been anticipated e.g. in Refs. \cite{Krumlinde1984, Rzaca2007, Winger2010}
while the contrary has been deduced by other authors, e.g. in Refs. \cite{Verney2007, Angelis2007, VandeWalle2009}.
One should however notice, that experimentally it is still not possible
to reach the $^{78}$Ni itself, and since the shell effects manifest themselves suddenly
at the shell closures, while being hindered by correlations in semi-magic nuclei,
it is not possible to conlude firmly on the rigidity of the $^{78}$Ni based on the currently
available data alone. To shed light on the physics of this nucleus one needs a theory, 
capable to reproduce the spectroscopic details of the region,
and robust enough to be extrapolated to the yet unknown. 
As the crucial information  
about the underlying shell evolution can be extracted from the structure
of odd nuclei, one is left with the choice of the
shell model to study the shell effects in the vicinity of $^{78}$Ni.

From the shell model point of view it is a very demanding region
but computionally tractable. Recently, we have addresed the problem of
the quenching of the $Z=28$ gap in $^{78}$Ni, using a large valence space
containing $pf$ orbitals for protons and $pf_{5/2}g_{9/2}$ orbitals for neutrons \cite{sieja-cu}.
These calculations allowed
to study in detail the role of proton core excitations
on the structure of the copper nuclei ($Z=29$) and proved, that
these degrees of freedom are crucial for a correct description of nuclear
data in this region. This appeared due to a slight reduction
of the $Z=28$ gap, of about 0.7MeV between $^{68}$Ni and $^{78}$Ni.

In this work we discuss the evolution of the neutron $N=50$
gap due to the proton-neutron interaction between $^{78}$Ni and $^{86}$Kr
and due to the neutron-neutron interaction between $^{68}$Ni and $^{78}$Ni.
 The calculations are performed in an enlarged
model space, which contains $pf$-shell orbitals for protons and $f_{5/2},p, g_{9/2},d_{5/2}$
orbitals for neutrons. The effective interaction starts with the same sets of two-body matrix elements
as used in Ref. \cite{Lenzi2010} however we have introduced further monopole changes to
constrain the proton gap evolution from $^{68}$Ni to $^{78}$Ni, deduced in Ref. \cite{sieja-cu},
as well as fine changes in the multipole part of the interaction, which
however leave unchanged the physics of the {\it island of inversion} studied in
\cite{Lenzi2010}. To probe reliability of such an interaction we investigate the low lying states of
even-even $N=50$ nuclei, between $^{78}$Ni and $^{84}$Se as well as in even-odd
 $N=49$ isotones, between $^{79}$Zn to $^{85}$Kr, which provides the insight into the
evolution of the $N=50$ gap. We have dwelled on the evolution of the 
$Z=28$ gap and its consequences on the calculated magnetic moments of copper isotopes 
in a previous work \cite{sieja-cu} and recently also calculations of magnetic moments in zinc isotopes
\cite{Georgi2011} became available. For the completeness of SM description in this region, 
we discuss here transition rates in nickel and copper chains.

The calculations are performed using the m-scheme code {\small ANTOINE} \cite{ANTOINE}, allowing
up to 12p-12h excitations with respect to $\pi 0f_{7/2}$ and $\nu 0g_{9/2}$
orbitals. The largest dimensions of matrices treated here reach $3\cdot10^9$.

\begin{figure}
\begin{center}
\includegraphics[scale=0.35]{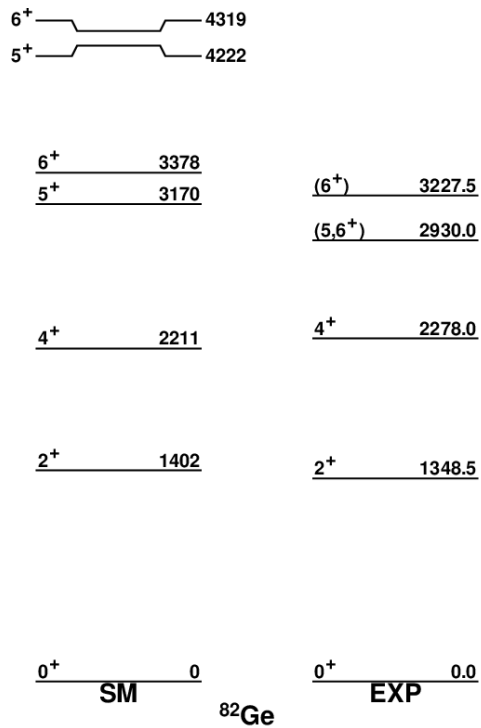}
\caption{Calculated vs experimental spectrum of $^{82}$Ge.\label{fig-ge}}
\end{center}
\end{figure}

Let us start the discussion with reminding the current status of experimental knowledge 
on the structure of $N=49$ and $N=50$ isotones.
In the low lying spectra of the $N=49$ nuclei one observes the negative parity
states, corresponding to the holes in the $p_{1/2}$ orbital, as well
as positive parity states among which the lowest excited $5/2^+$ may correspond to
a 1p-1h excitation across the $N=50$ gap and can be thus addressed in our model
space which includes the $d_{5/2}$ neutron orbital. The systematics of $5/2^+$
states in $N=49$ isotones is known up to $^{81}$Ge (4 protons away from $^{79}$Ni)
but experimental data should be soon available as far as for $^{79}$Zn (2 protons away from $^{79}$Ni) \cite{Orlandi}.
The states corresponding to the excitations across the $N=50$
gap serve as a perfect test for its size
assumed in our interaction in the vicinity of $^{78}$Ni.
In the spectra of even-even nuclei, 1p-1h excitation can form the first excited $5^+, 6^+$
states. They are known experimentally up to $^{82}$Ge 
and the relatively low excitation energy of these states at $Z=32$ has lead the authors
of Ref. \cite{Rzaca2007} to anticipate the possible
weakening of the $N=50$ closure. We start thus the presentation of our results with the example of 
$^{82}$Ge. In Figure \ref{fig-ge} we show the spectra calculated in our approach compared
to experimental data for the case of $^{82}$Ge. The known energy levels $2^+, 4^+$ are reproduced with a great accuracy
and theoretical counterparts for the experimental candidates of $5^+, 6^+$ spins are located
within 200~keV.  We have added to the spectrum also the second excited
$5^+$ and $6^+$ levels. The occupation numbers in the corresponding wave functions are given in Table \ref{tab-wf}.
The first excited $2^+, 4^+$ states are clearly of a proton nature.
On the contrary, it is seen that the lowest calculated $5^+, 6^+$ levels correspond indeed to the 1p-1h
excitations to the $\nu d_{5/2}$ orbital, as proposed in Ref. \cite{Rzaca2007}, and thus probe the size of the $N=50$ gap 
assumed in our model at $Z=32$. The second calculated $5^+, 6^+$ states,
located at around 4~MeV, are again of the proton nature, having nearly the same occupations
of the neutron $g_{9/2}, d_{5/2}$ orbitals as the ground state.

\begin{table}
\begin{center}
\caption{Occupation numbers in the wave function of low lying excited states
in $^{82}$Ge. \label{tab-wf}}
\begin{tabular}{c|ccc|cc}
\hline
\hline
& \multicolumn{3}{c|}{protons} &\multicolumn{2}{c}{neutrons}\\
 $J^\pi$ & $f_{7/2}$ & $f_{5/2}$ & $p$ & $g_{9/2}$ &$d_{5/2}$\\
\hline
    $0^+$   &7.64  &3.74  &0.41 &9.65&0.37\\
    $2^+$   &7.77  &3.73  &0.33 &9.60&0.42\\
    $4^+$   &7.84  &3.79  &0.23 &9.6&0.36\\
    $5_1^+$ &7.59  &3.61  &0.50 &8.53&1.48\\
    $6_1^+$ &7.59  &3.62  &0.49 &8.57&1.44\\
    $5_2^+$ &7.79  &3.01  &0.81 &9.52&0.49\\   
    $6_2^+$ &7.82  &3.04  &1.02 &9.64& 0.37\\  
\hline
\hline
\end{tabular}
\end{center}
\end{table}

\begin{figure}
\begin{center}
\includegraphics[scale=0.4]{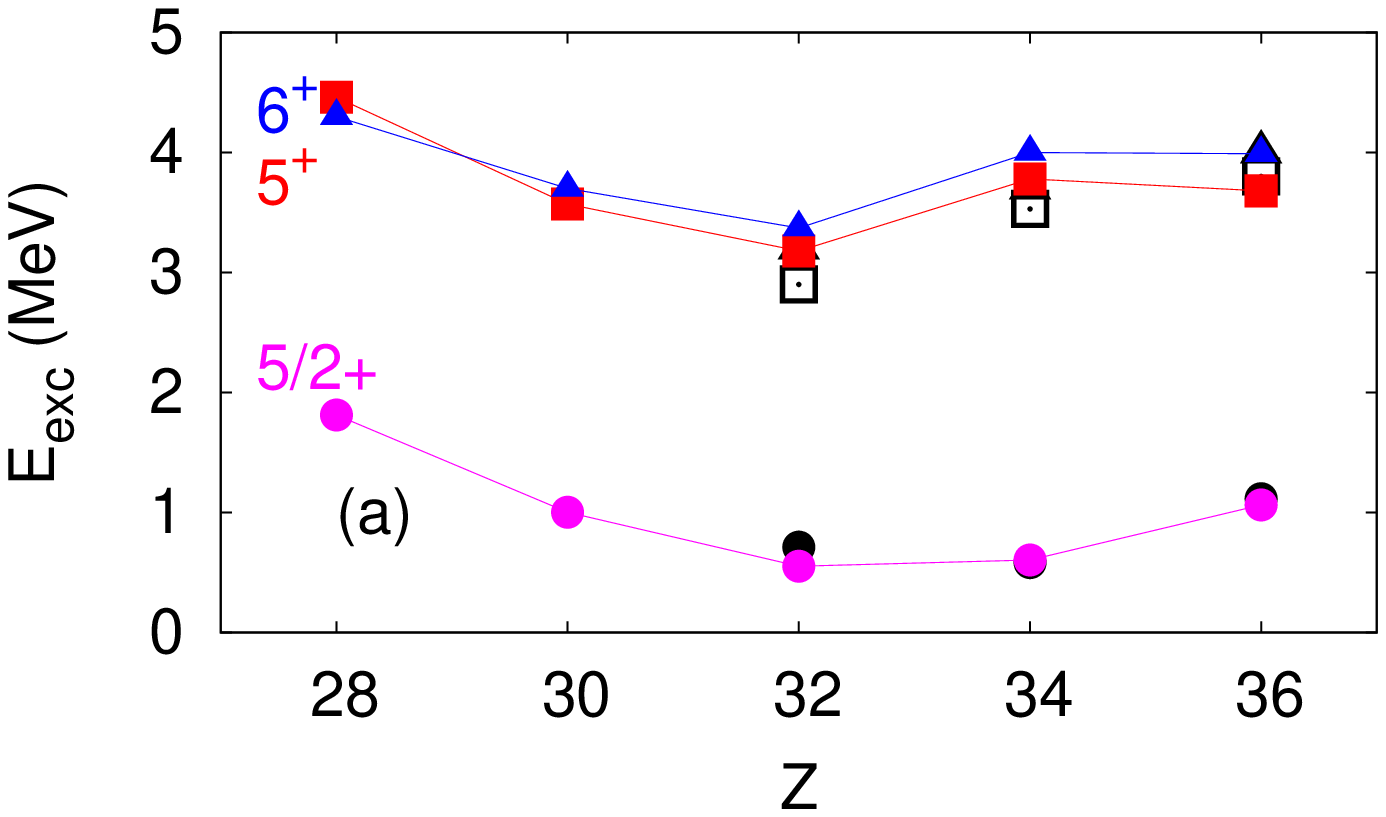}\\
\includegraphics[scale=0.4]{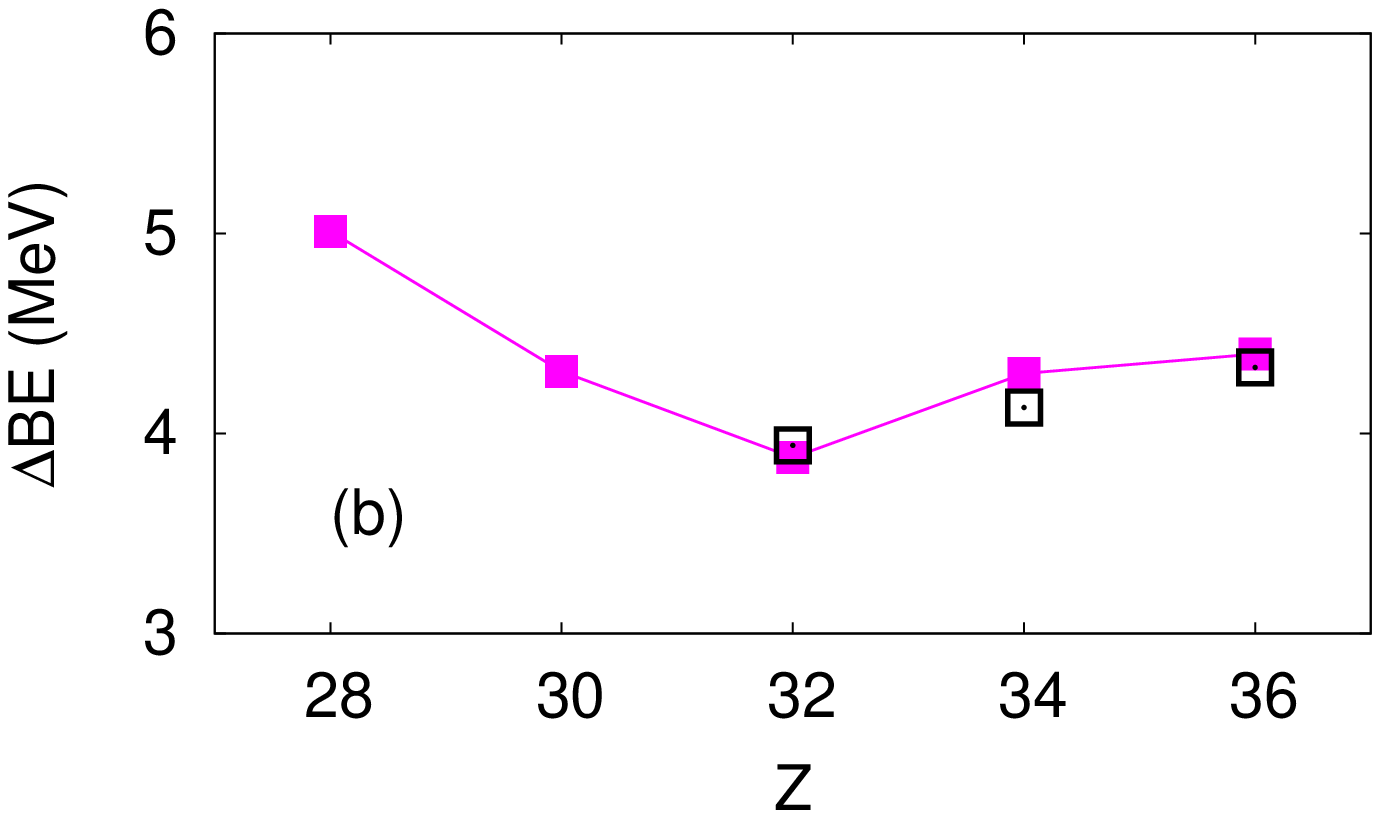}
\caption{Systematics of the low lying 1p-1h states in $N=50$
and $N=49$ isotones (a) and the evolution of the $N=50$ 
gap calculated from masses (b). Black symbols represent the experimental data.
Extrapolated values from \cite{AME2003}
have been used for $^{79,81}$Zn and $^{83}$Ge in the gap estimates.\label{fig-sys}  }
\end{center}
\end{figure}

In Fig. \ref{fig-sys} we present the systematics of lowest excited $5^+,6^+$
states from $Z=28$ to $Z=36$ compared to the known experimental data, which 
probes the evolution of the gap when moving away from $Z=32$.
Our calculations match well the decreasing trend observed from Kr to Ge, 
but the increase of the excitation energy is predicted towards Ni.
In addition, we present in the same figure the systematics of the calculated
and experimentally known $5/2^+$ states in $N=49$ isotones. 
These states are correctly reproduced and, similarily to the case of even-even nuclei,
the increase of the $5/2^+$ is observed towards Ni. The first excited state $5/2^+$ in $^{79}$Zn
is predicted to be located at 1~MeV.

\begin{figure}
\begin{center}
\includegraphics[scale=0.4]{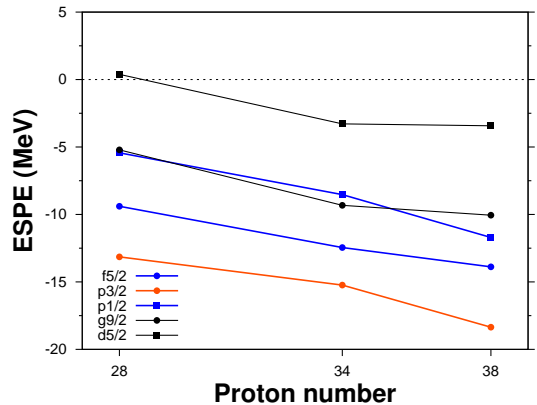}
\caption{Evolution of the neutron effective single particle energies
with the  proton number at $N=50$. \label{fig-espe}}
\end{center}
\end{figure}

In both systematics a minimum appears around $Z=32$. Such a minimum
however does not reflect any changes in the spherical mean-field in these nuclei.
This is illustrated in Figs. \ref{fig-sys} and \ref{fig-espe}.
The former shows the correlated gap, i.e. the gap obtained from mass differences:
\begin{equation}
\Delta=BE(Z+1,N)+BE(Z-1,N)-2BE(Z,N).
\end{equation}
The latter is the shell model prediction for the evolution of the
effective single particle energies (ESPE), which are obtained as the differences 
of monopole energies of closed shell and closed shell $\pm1$ particle configurations.
The difference of the corresponding ESPE defines the uncorrelated
shell gap and allows to separate the spherical mean-field from correlation effects
contained in $\Delta$. One should also note that the ESPE represent schematically the evolution of the monopole field 
in nuclei, which is not an observable itself. In particular, it assumes a given filling ordering 
scheme broken in reality by the residual interaction but for many years the monopole Hamiltonian has been used succesfully 
to investigate and visualize the shell effects in nuclei.

In Fig. \ref{fig-espe} we present the evolution of the neutron effective single particle 
energies (ESPE) with the proton number, where no clear variation of the gap is observed
between $^{78}$Ni and $^{86}$Kr. At the same time,
the gap calculated from mass differences varies from 5~MeV in $^{78}$Ni to
only 3.9~MeV in $^{82}$Ge and then increases again towards $^{86}$Kr as can be seen from
Fig. \ref{fig-sys}. This indicates that
the correlation effects lead to a minimum in the mass gap at $Z=32$, which explains the
observed pattern of the excited states, but at the same time,
no indication for the weakening of the spherical gap towards $^{78}$Ni is found. 
One should point out, that such a parabolic behavior is typical for  
semi-magic nuclei, where the correlations are the largest in the mid-shell 
(we can quote the scandium chain as a prominent example).

Let us now discuss the variation of the $N=50$ gap from $^{68}$Ni
to $^{78}$Ni due to the $T=1$ part of the nuclear force.
It has been anticipated already decades ago, that the pure 2-body interactions
are not sufficient to reproduce the shell evolution, and in particular the
observed spin-orbit splittings. The blame has been put on the missing three-body force \cite{Zuker2003}.
Only recently the first attempts of SM calculations with effective 3-body forces
based on the chiral N3LO potential  
became available \cite{Holt1, Holt2}. These calculations suggest that at least some of
the missing repulsion between $d_{5/2}$ and $s_{1/2}$ orbitals in oxygen and
$f_{7/2}$ and $p_{3/2}$ orbitals in calcium can
be gained by including the 3-body contribution between 2 valence and one core particles.
As the 3-body effects can explain the spin-orbit shell closures observed experimentally
at $Z=8$ ($N=1$ harmonic oscillator shell closure)
and $Z=20$ ($N=2$ h.o. closure), one could search for similar effects at $Z=40$ ($N=3$).
However, the harmonic oscillator to spin-orbit closure
transition is taking place between $Z=20$ and $Z=40$, making the next place in the Segr\`e chart
available for such studies already between $^{68}$Ni and $^{78}$Ni.

\begin{figure}
\begin{center}
\includegraphics[scale=0.4]{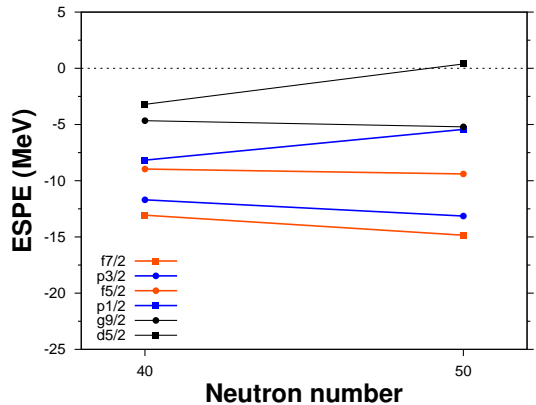}
\caption{Evolution of the neutron effective single particle energies
between $^{68}$Ni and $^{78}$Ni. \label{fig-nespe} }
\end{center}
\end{figure}

In Fig. \ref{fig-nespe} we show the evolution of
neutron effective single particle energies in the considered region.
Apparently, in the nickel chain a shell gap is created when adding neutrons
to the $g_{9/2}$ orbital due to the repulsive $T=1$ $g_{9/2}$-$d_{5/2}$
monopole interaction, in full analogy to what is observed in oxygens and calciums.
It appears that the nickel chain offers indeed another possibility for investigating the 3-body effects
in medium mass nuclei. Note, that such a behavior of $N=50$ gap has not been imposed in our model but results
from the study of the island of inversion around $^{64}$Cr, which allows to constrain approximately the 
position of the $d_{5/2}$ orbital at $N=40$ and from the study of the $N=49-50$ isotones, 
which allowed for testing the size of the $N=50$ gap in $^{78}$Ni.


\begin{figure}
\begin{center} 
\includegraphics[scale=0.4]{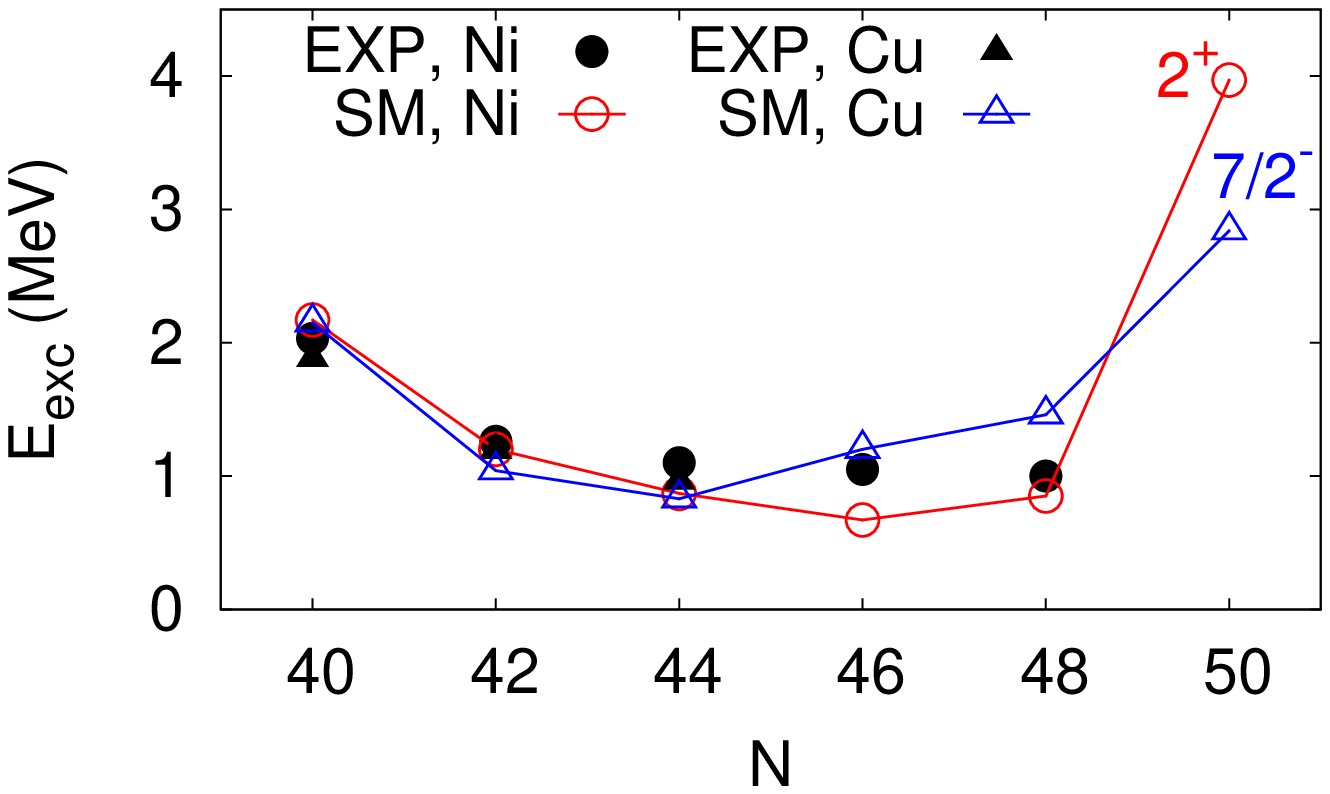}\\
\includegraphics[scale=0.4]{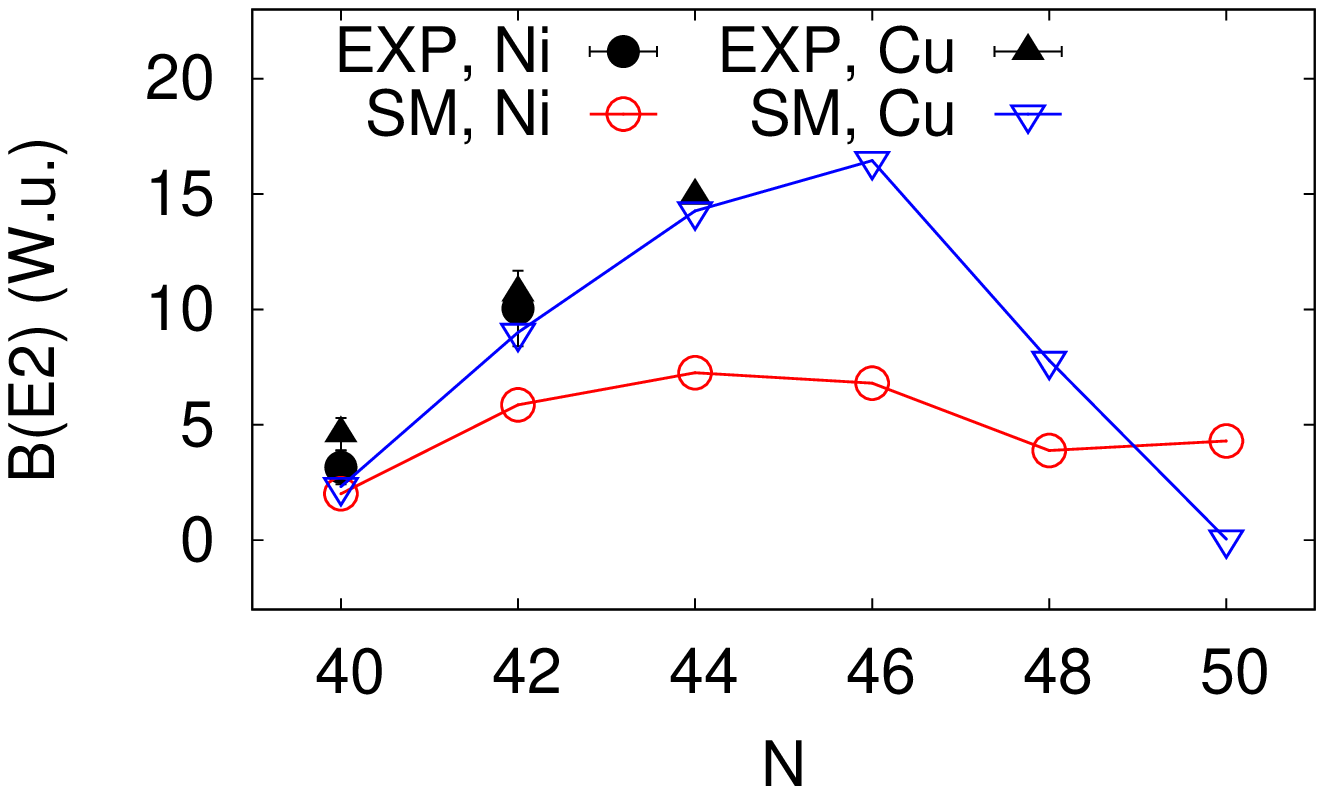}
\caption{Excitation energies and reduced transition probabilities in the nickel and copper chains.
The $B(E2; 2^+\rightarrow 0^+)$ is plotted for Ni, $B(E2; 7/2^-\rightarrow 3/2^-)$
for coppers. \label{fig-nickel}}
\end{center}
\end{figure}

In a previous work \cite{sieja-cu} we have belabored on the evolution of the $Z=28$
gap, its weaknening between $^{68}$Ni and $^{78}$Ni and the consequences 
on the shell model description of this region, which we have illustrated with the systematics 
of magnetic moments in copper isotopes. These calculations have been achieved in a smaller model space, i.e.
without the $d_{5/2}$ orbital. In the isotopic chains near $Z=28$ (Cu, Ni, Zn)
the major effect of the inclusion of the $d_{5/2}$ orbital, the quadrupole partner of $g_{9/2}$, is enhancing the calculated $B(E2)$ values
by $10-20\%$.
Recently we have discussed the transition values and magnetic moments of zinc isotopes
\cite{Georgi2011} obtained in such an enlarged model space, finding a fair agreement with experimental data
for both quantities. Also transition rates in odd-odd coppers have been addressed within 
this model space \cite{Elisa2011}, showing 
how the inclusion of simultaneous excitations across the $Z=28$ and $N=50$ 
gaps can account for the large core polarization observed in SM calculations within truncated model spaces
outside the $^{56}$Ni core. 
Here, to complete the SM description of this region, we present the systematics of $2^+$ states
and $B(E2; 2^+\rightarrow 0^+)$ values in nickel isotopes in Figure \ref{fig-nickel}. We have added for comparison the 
systematics of the $7/2^-$ states and  $7/2^-\rightarrow 3/2^-$ 
transition rates in copper isotopes.
The $7/2^-$ state in coppers may be formed by a coupling of the odd proton 
to the $2^+$ state of the nickel core, thus the transition rates might follow the trend of the 
$B(E2; 2^+\rightarrow 0^+)$ ones in nickels. The wave functions of the $2^+$ and $7/2^-, 3/2^-$ states 
are given in Table \ref{tab-wf1}. 

It is seen that at shell closures $N=40$ and $N=50$
the $7/2^-$ state corresponds to one hole in the $\pi f_{7/2}$ orbital, otherwise it has
a more collective nature, with the odd proton distributed between $f_{5/2}$
and $p$ orbitals. One should also remind, that the $3/2^-$ state has a single particle nature
at $N=40$ and then gains in collectivity due to the rapidly descending $f_{5/2}$ orbital
which becomes the ground state in $^{75}$Cu. As seen from the Table, it contains again $\sim1$ particle
in the $p_{3/2}$ orbital when the $N=50$ closure is reached.

\begin{table}
\begin{center}
\caption{Occupation numbers in the $2^+$ states of nickel isotopes ($Z=28$) and the lowest excited $7/2^-$
states in copper isotopes ($Z=29$). \label{tab-wf1}}
\begin{tabular}{cl|cccc|cc}
\hline
\hline
 & & \multicolumn{4}{c|}{protons} &\multicolumn{2}{c}{neutrons}\\
 $N$& $J^{\pi}$  & $f_{7/2}$ & $f_{5/2}$ & $p_{3/2}$ & $p_{1/2}$ & $g_{9/2}$ &$d_{5/2}$\\
\hline
  $40$ &  $2^+$   &    7.25 & 0.24 & 0.45& 0.06 & 2.20 & 0.13  \\
       &  $7/2^-$ &    6.49 & 1.26 & 0.77& 0.46& 3.24 & 0.52  \\
       &  $3/2^-$ &    7.80 & 0.13 & 1.04& 0.03 & 1.80 & 0.60\\
\hline  
  $42$ &  $2^+$   &    7.56 & 0.20 & 0.21 & 0.03 & 2.53 & 0.28  \\
       &  $7/2^-$ &    7.44 & 0.63 & 0.73 & 0.19 & 3.17 & 0.38  \\
       &  $3/2^-$ &    7.58 & 0.36 & 0.92 & 0.14 & 2.82 & 0.32 \\
\hline
  $44$ &  $2^+$   &    7.49 & 0.25 & 0.21 & 0.05 & 3.94 & 0.58\\
       &  $7/2^-$ &    7.64 & 0.59 & 0.57 & 0.20 & 3.72 & 0.53\\ 
       &  $3/2^-$ &    7.64 & 0.55 & 0.59 & 0.22 & 3.91 & 0.54\\  
\hline
  $46$ &  $2^+$   &    7.49 & 0.29 & 0.18 & 0.04 & 5.60 & 0.59\\
       &  $7/2^-$ &    7.59 & 0.76 & 0.43 & 0.22 & 5.50 & 0.60\\
       &  $3/2^-$ &    7.59 & 0.75 & 0.40 & 0.25 & 5.53 & 0.61 \\
\hline
  $48$ &  $2^+$   &    7.58 & 0.27 & 0.13 & 0.02 & 7.65 & 0.40 \\
       &  $7/2^-$ &    7.64 & 1.07 & 0.22 & 0.07 & 7.59 & 0.45 \\
       &  $3/2^-$ &    7.64 & 1.02 & 0.22 & 0.11 & 7.61 & 0.44 \\
\hline
  $50$ &  $2^+$   &    7.47 & 0.35 & 0.15 & 0.03 & 8.67 & 1.35 \\
       &  $7/2^-$ &    6.74 & 2.03 & 0.16 & 0.07 & 9.63 & 0.39 \\
       &  $3/2^-$ &    7.64 & 0.37 & 0.94 & 0.04 & 9.76 & 0.25\\
\hline
\hline
\end{tabular}
\end{center}
\end{table}

As has been shown in \cite{Dijon2011}, our calculations account very well for the complex physics of $^{68}$Ni, in which 
the coexistence of spherical and deformed $0^+$ states has been found experimentally.
This evidences that the interplay between the size of the $Z=28$ and $N=40$ gaps constrained in the monopole Hamiltonian
and the correlations brought to the system by the configuration mixing 
should be reliable in the shell model interaction. As can be seen from Figure \ref{fig-nickel}, the evolution
of $2^+$ and $7/2^-$ energies away from $N=40$ is correct in the present calculations, too.
The rapid increase of collectivity between $^{68}$Ni and $^{70}$Ni, reflected in the enhancement of $B(E2; 2^+\rightarrow 0^+)$ value, 
is however underestimated considerably in $^{70}$Ni (note that a larger value
for this transition was found with the original LNPS interaction \cite{Lenzi2010}, which may be considered as a theoretical
uncertainty due to the slight variation of the parameters of the Hamiltonian). 
Interestingly, the known $B(E2; 7/2^-\rightarrow 3/2^-)$ transition in copper isotopes is accurately reproduced
also at $N=42$. This is however as well the only case between $N=42-48$, where
the neutron occupancies are clearly different in the $2^+$ of Ni and the first excited $7/2^-$ level of Cu, which may
point to a slightly inaccurate balance between proton and neutron excitations in the theoretical model.
More experimental information on the systematics of $B(E2; 2^+\rightarrow 0^+)$ would be helpful to
understand the puzzling case of $^{70}$Ni. 

The ground state of $^{78}$Ni is calculated to have
$79\%$ of the closed shell configuration,
the largest value in the whole nickel chain
(around $60\%$ of closed shell is found in $pf$ calculations in $^{56}$Ni with various interactions and only $50\%$
in $^{68}$Ni, here or in previous SM studies \cite{Sorlin2002}).   
The first $2^+$ state in $^{78}$Ni is predicted at nearly 4~MeV, a value analogous to 
the ${2^+}$ state of its "big sister" $^{132}$Sn. This first excited state 
is of a neutron nature, having 1.35
particle in the $d_{5/2}$ orbital (0.16 particle in the $0^+$ state).
The gaps estimated from calculated mass differences in $^{78}$Ni are 5~MeV for neutrons and 
5.25~MeV for protons.    
   
In summary, we have performed large scale shell model calculations in the 
$\pi(fp)$-$\nu(fpgd)$ model space, in the vicinity of $^{78}$Ni. Using phenomenological 
interaction whose monopole drifts allow to reproduce the known experimental data
in the island of inversion region and in $N=40-50$ nuclei,
we have studied the evolution of the $N=50$ shell gap due to the proton-neutron
interaction between $^{86}$Kr and $^{78}$Ni. The calculations point to a minimum
of the mass gap in $^{82}$Ge and its increase towards $^{78}$Ni. We predict the location
of the first excited $5/2^+$ in $^{79}$Zn around 1~MeV and at nearly 2~MeV in $^{78}$Ni itself.
The evolution of the $N=50$ gap along the nikel chain bears similarity to what has been
attributed to the action of the effective 3-body forces in oxygen and calcium chains.
A study analogous to Refs. \cite{Holt1, Holt2} is called for to examine further the origin of the
$N=50$ shell gap in $^{78}$Ni. 
The systematics of excited states and $B(E2)$ values in Ni chain has been presented, 
which purports a high lying first excited $2^+$ state in $^{78}$Ni.
We conclude on the robustness of both, $Z=28$ and $N=50$ gaps from the spectroscopic study
in this region. Future experiments with a new generation of facilities 
are necessary to confirm our predictions for the 
shell closures in $^{78}$Ni.

\bibliographystyle{apsrev}

\end{document}